# Non-perturbative O(a) improvement of lattice QCD

Martin Lüscher[a], Stefan Sint[b], Rainer Sommer[c,d], Peter Weisz[e] and Ulli Wolff[f]

[a] Deutsches Elektronen-Synchrotron DESY,
   Notkestrasse 85, D-22603 Hamburg, Germany

[b] SCRI, The Florida State University, Tallahassee, FL 32306-4052, USA

[c] DESY-IfH Zeuthen, Platanenallee 6, D-15738 Zeuthen, Germany

[d] CERN, Theory Division, CH-1211 Genève 23, Switzerland

[e] Max-Planck-Institut für Physik,
   Föhringer Ring 6, D-80805 München, Germany

[f] Humboldt Universität, Institut für Physik,
   Invalidenstrasse 110, D-10099 Berlin, Germany

## Abstract

The coefficients multiplying the counterterms required for O($a$) improvement of the action and the isovector axial current in lattice QCD are computed non-perturbatively, in the quenched approximation and for bare gauge couplings $g_0$ in the range $0 \leq g_0 \leq 1$. A finite-size method based on the Schrödinger functional is employed, which enables us to perform all calculations at zero or nearly zero quark mass. As a by-product the critical hopping parameter $\kappa_c$ is obtained at all couplings considered.

September 1996

hep-lat/9609035 v2   29 Oct 1996

# 1. Introduction

Numerical simulations of lattice QCD are limited to lattices with lattice spacings $a$ that are often not very much smaller than the relevant physical scales. The associated systematic errors can be rather large and must be studied carefully. It has been known for a long time that the lattice effects can be reduced by choosing an improved discretization of the continuum theory. The subject has recently found renewed interest and substantial progress has been made in various directions (see ref. [1] for a review and an up-to-date list of references). Here we consider on-shell $\mathrm{O}(a)$ improved lattice QCD, where the improvement is achieved by adding a few higher-dimensional local counterterms to the lattice action and the composite fields of interest.

In Wilson's original formulation of lattice QCD the leading cutoff effects in physical amplitudes are proportional to $a$. As explained in refs. [2,3] the presence of these terms is easily seen by studying the conservation of the isovector axial current in suitable correlation functions. Moreover, it has been noted that the coefficients of the counterterms required for $\mathrm{O}(a)$ improvement of the action and the current, $c_{\mathrm{sw}}$ and $c_{\mathrm{A}}$, can be determined by imposing the validity of the PCAC relation up to corrections of order $a^2$. The idea has been shown to work out in perturbation theory [4] and we now apply it in quenched QCD to compute $c_{\mathrm{sw}}$ and $c_{\mathrm{A}}$ non-perturbatively using numerical simulations.

The precise formulation of the improved theory and the theoretical framework that goes along with it will not be reviewed here. Instead we assume that the reader is familiar with ref. [3], where the basic definitions are all given explicitly. The notations introduced there are taken over completely without further notice. Equations in ref. [3] are referred to by prefixing a Roman "I" to the equation number.

In sect. 2 we introduce the correlation functions to be studied and then proceed to describe a few details of the numerical simulations that we have performed (sect. 3). A technical problem having to do with the occurrence of quark zero-modes and the fundamental limitations of the quenched approximation is discussed in sect. 4. In the main part of the paper, sects. 5 and 6, we explain the computation of $c_{\mathrm{sw}}$ and $c_{\mathrm{A}}$ and present our results. With little additional effort the critical hopping parameter $\kappa_{\mathrm{c}}$ can also be calculated, for any value of the bare coupling $g_0$ between 0 and 1 (sect. 7). A few concluding remarks are collected in sect. 8.

## 2. Correlation functions

Following refs. [2,3] the lattice corrections to the PCAC relation will be studied in the framework of the Schrödinger functional. The precise definition of the latter is given in sects. 4 and 5 of ref. [3]. In particular, the form of the $O(a)$ boundary counterterms that must be included in the action, when the goal is to improve the Schrödinger functional itself, has been derived there.

For the present investigation these counterterms are not required, because they only affect the PCAC relation at order $a^2$ (cf. subsect. 6.1 of ref. [3]). We thus omit them in the following except for the second term in eq. (I.5.6), with $c_t$ given by the one-loop expression $1 - 0.089 \times g_0^2$. This enables us to obtain some direct checks on the current version of the simulation program by comparing with simulation data generated earlier in the course of a calculation of the running coupling in the pure gauge theory [5], where the term had been included for good reason.

In the following the time-like extent $T$ of the lattice is always taken to be twice the spatial size $L$. The gauge group is $SU(3)$ and the boundary values of the gauge field, $C$ and $C'$, are assumed to be constant diagonal matrices as in subsect. 6.2 of ref. [3]. We shall be interested in the correlation functions

$$f_{\mathrm{A}}(x_0) = -a^6 \sum_{\mathbf{y},\mathbf{z}} \tfrac{1}{3} \langle A_0^a(x)\, \bar{\zeta}(\mathbf{y})\gamma_5 \tfrac{1}{2}\tau^a \zeta(\mathbf{z}) \rangle, \tag{2.1}$$

$$f_{\mathrm{P}}(x_0) = -a^6 \sum_{\mathbf{y},\mathbf{z}} \tfrac{1}{3} \langle P^a(x)\, \bar{\zeta}(\mathbf{y})\gamma_5 \tfrac{1}{2}\tau^a \zeta(\mathbf{z}) \rangle, \tag{2.2}$$

which involve the boundary quark fields $\zeta$ and $\bar{\zeta}$ at time $x_0 = 0$. For the (unimproved, unrenormalized) axial current and density the local expressions

$$A_\mu^a(x) = \overline{\psi}(x)\gamma_\mu\gamma_5 \tfrac{1}{2}\tau^a\psi(x), \tag{2.3}$$

$$P^a(x) = \overline{\psi}(x)\gamma_5 \tfrac{1}{2}\tau^a\psi(x), \tag{2.4}$$

are employed, where $\tau^a$ is a Pauli matrix acting on the flavour indices of the quark field $\psi(x)$.

A second set of correlation functions, $f'_{\mathrm{A}}$ and $f'_{\mathrm{P}}$, is defined through

$$f'_{\mathrm{A}}(T - x_0) = +a^6 \sum_{\mathbf{y},\mathbf{z}} \tfrac{1}{3} \langle A_0^a(x)\, \bar{\zeta}'(\mathbf{y})\gamma_5 \tfrac{1}{2}\tau^a \zeta'(\mathbf{z}) \rangle, \tag{2.5}$$

$$f'_{\mathrm{P}}(T-x_0) = -a^6 \sum_{\mathbf{y},\mathbf{z}} \tfrac{1}{3} \langle P^a(x)\,\bar{\zeta}'(\mathbf{y})\gamma_5 \tfrac{1}{2}\tau^a \zeta'(\mathbf{z})\rangle. \tag{2.6}$$

Here the axial current and density are probed by the boundary quark fields at time $T$. Note that $f_{\mathrm{X}}$ and $f'_{\mathrm{X}}$ are related to each other through a time reflection. Under this transformation the boundary values $C$ and $C'$ are interchanged so that $f_{\mathrm{X}}$ and $f'_{\mathrm{X}}$ are not the same in general.

As usual the integration over the quark fields is carried out analytically before the numerical simulation is set up. The boundary conditions satisfied by the quark fields have to be taken into account in this step. This is discussed in detail in sect. 2 of ref. [4]. The outcome is that any correlation function of the bulk and boundary quark fields, in any given background gauge field, can be calculated by applying Wick's theorem with the appropriate two-point contractions.

In the case of the correlation functions $f_{\mathrm{A}}$ and $f_{\mathrm{P}}$ there is only one way to contract the quark fields and one ends up with

$$f_{\mathrm{X}}(x_0) = \tfrac{1}{2}\left\langle \operatorname{tr}\left\{ H(x)^\dagger \Gamma_{\mathrm{X}} H(x)\right\}\right\rangle_{\mathrm{G}}, \tag{2.7}$$

where $\Gamma_{\mathrm{A}} = -\gamma_0$ and $\Gamma_{\mathrm{P}} = 1$. The matrix $H(x)$, defined below, is the quark propagator from the boundary at time $0$ to the point $x$ in the interior of the space-time volume. In the quenched approximation the expectation value $\langle\ldots\rangle_{\mathrm{G}}$ is to be taken in the pure gauge theory. A similar expression is obtained for the other correlation functions $f'_{\mathrm{A}}$ and $f'_{\mathrm{P}}$.

The matrix $H(x)$ has colour and Dirac indices. It is defined through

$$(D + \delta D + m_0)H(x) = 0, \qquad 0 < x_0 < T, \tag{2.8}$$

and the boundary conditions

$$P_+ H(x)|_{x_0=0} = P_+, \qquad P_- H(x)|_{x_0=T} = 0. \tag{2.9}$$

In this equation $D$ denotes the Wilson-Dirac operator, eq. (I.2.3), and $\delta D$ derives from the O($a$) counterterms in the improved quark action. Since the quark boundary counterterms have been dropped, only the Sheikholeslami-Wohlert term contributes and

$$\delta D \psi(x) = c_{\mathrm{sw}} \tfrac{i}{4} a \sigma_{\mu\nu} \widehat{F}_{\mu\nu}(x) \psi(x). \tag{2.10}$$

The numerical solution of eq. (2.8) is discussed in subsect. 3.2.

# 3. Numerical simulation

The numerical simulations reported in this paper have been performed on an APE/Quadrics computer with 256 nodes. This machine has a SIMD architecture, which proves to be well suited for the simulation of lattice QCD with $O(a)$ improvement and Schrödinger functional boundary conditions. Following common usage we shall now often quote values of $\beta = 6/g_0^2$ and $\kappa = (8 + 2am_0)^{-1}$ instead of the bare coupling and mass.

## 3.1 Simulation algorithm

To generate a representative ensemble of gauge fields, a hybrid over-relaxation algorithm was used [6,7]. The local updates were ordered according to the "SF-scheme" described in subsect. 3.2 of ref. [8]. Both micro-canonical reflections and heatbath steps [9,10] are performed for three different embedded SU(2) subgroups of SU(3). The efficiency of the algorithm depends on these details and also on the number $N_{\mathrm{OR}}$ of micro-canonical (or over-relaxation) sweeps per heatbath sweep, which was taken to be 5 in most cases. In the following the term "iteration" denotes a complete update cycle of one heatbath sweep followed by $N_{\mathrm{OR}}$ micro-canonical sweeps.

An interesting observable constructed from the Schrödinger functional is the renormalized coupling studied in refs. [5,8]. The coupling can be computed at almost no cost and thus provides a good opportunity to monitor the dynamics of the simulation. In most cases we had several $10^5$ gauge configurations and the integrated autocorrelation times of the coupling could be estimated reliably. They range from a third of an iteration to two iterations, the latter being reached at low values of $\beta$.

Taking this into account we decided to evaluate the correlation functions $f_{\mathrm{A}}$ etc. by averaging over sequences of gauge field configurations separated by 50 iterations. The individual "measurements" of the correlation functions are then expected to be statistically independent for all practical purposes (cf. subsect. 3.3).

## 3.2 Solving Dirac's equation

We now proceed to discuss the solution of the boundary value problem defined through eqs. (2.8) and (2.9). If we introduce the modified matrix

$$\widetilde{H}(x) = H(x) - \delta_{x_0 0} P_+, \tag{3.1}$$

it is straightforward to show that

$$(D + \delta D + m_0)\widetilde{H}(x) = \frac{1}{a}\delta_{x_0 a}U(x - a\hat{0}, 0)^{-1}P_+, \qquad 0 < x_0 < T. \qquad (3.2)$$

Since $\widetilde{H}(x)$ vanishes at $x_0 = 0$ and $x_0 = T$, this is just a system of linear equations which can be solved using standard methods [11]. Note that $\widetilde{H}(x)$ is a complex $12 \times 12$ matrix at each point $x$. One thus has to solve 12 systems of vector equations, but in view of the projector $P_+$ on the right-hand side of eq. (3.2) only half of them need to be considered.

To solve the linear equations we employed the stabilized biconjugate gradient algorithm (BiCGstab) with even-odd preconditioning [12,13]. As has previously been observed, the implementation of this sort of preconditioning does not present any great difficulties in the $O(a)$ improved theory [14]. Before starting the iteration one only has to invert the hermitean matrix

$$1 + \kappa c_{\mathrm{sw}}\frac{i}{2}a^2\sigma_{\mu\nu}\widehat{F}_{\mu\nu}(x) \qquad (3.3)$$

on the even sites $x$ of the lattice. If one employs a chiral representation of the Dirac matrices, as in appendix A of ref. [3], the matrix assumes a block-diagonal form. The inversion of the two $6 \times 6$ blocks on the diagonal is then achieved by applying a Householder triangularization with subsequent backward substitution [15]. We found this algorithm particularly suitable for a SIMD machine, where strategies that require pivotization are hard to implement efficiently. There is no rigorous inequality excluding the singularity of the $6 \times 6$ blocks for completely arbitrary gauge fields, but we have never encountered any problem with this in the range of $\kappa$ and $c_{\mathrm{sw}}$ considered.

The BiCGstab iterations were stopped when the norm of the residue vector was smaller than $\epsilon$ times the norm of the solution vector of the even-odd preconditioned system. Varying $\epsilon^2$ from $10^{-11}$ to $10^{-13}$, we found that the ratios of correlation functions we are ultimately interested in changed by no more than $5 \times 10^{-5}$, which is below the statistical precision of our calculation. We then adopted $\epsilon^2 = 10^{-13}$ as the stopping criterion in all production runs.

On the $16 \times 8^3$ lattice, and for all quark masses considered, around 100 iterations were needed to reach convergence, while on the $32 \times 16^3$ lattices this number is generally twice as large. This is the expected behaviour in physically small volumes, where perturbation theory may be used to show that the spectrum of the Dirac operator has a gap of order $1/L$ [16,17]. Since the spectral radius of the operator is practically independent of the lattice size, the condition number of the system (3.2) grows linearly with $L$. The situation in physically large volumes (say $L \geq 1$ fm) is different and one finds that the number of iterations varies more appreciably with the quark mass and the gauge field: occasionally 300 iterations and more were needed to reach convergence.

Before starting the simulations we have checked that the BiCGstab algorithm is more efficient than the minimal residual and conjugate gradient algorithms [11], particularly for small and negative quark masses. In the most demanding case encountered ($L/a = 16$, $\beta = 6.0$, $am_{\mathrm{q}} = 0.02$ and non-perturbatively determined $c_{\mathrm{sw}}$) the average gain in CPU-time compared to the conjugate gradient algorithm was a factor of 3.

The correctness of our programs has been verified at small couplings $g_0$ by comparing with one-loop perturbation theory [4]. We have also written a set of Fortran-90 programs, which allows us to check the calculation of the correlation functions for individual gauge field configurations. In particular, the rounding errors associated with the 32 bit arithmetic on the APE computer could be shown to be completely negligible in our calculations.

### 3.3 Error analysis

We first remark that the statistical errors on the correlation functions can be reduced by averaging over the spatial coordinates $x_1, x_2, x_3$ in eq. (2.7). In general we are interested in calculating certain combinations of ratios of correlation functions and thus need to estimate the statistical errors associated with them. Moreover some of the computed quantities require an interpolation in the hopping parameter $\kappa$. In all cases the primary data are strongly correlated since they are obtained from the same set of gauge field configurations. This is taken care of by applying the jackknife method for the error estimation.

The calculations of $c_{\mathrm{sw}}$ and $c_{\mathrm{A}}$ presented in sects. 5 and 6 are based on ensembles of typically $32 \times 50$ configurations (the APE has been divided into 32 subsystems, each simulating an independent copy of the lattice). The integrated autocorrelation times have then been estimated by blocking the data before forming jackknife samples. No significant autocorrelations were found, in agreement with our experience with purely gluonic quantities (cf. subsect. 3.1).

For the computation of the critical hopping parameter a larger lattice is used and averages are taken over a much smaller ensemble of gauge field configurations. Typically we have of the order of 100 configurations and even less at small couplings $g_0$. In this case we could check for autocorrelations using block lengths 1 and 2 only. We found no significant difference in the errors. With such low statistics the errors have uncertainties of about 30%, but we did not care to increase the statistics, because the results are already rather precise even if we assume that the errors have been underestimated by a factor of 2.

## 4. Zero-modes and breakdown of the quenched approximation

In the course of our computations some very large statistical fluctuations have been observed at bare couplings $g_0 > 1$ and small quark masses. It is for this reason that we shall restrict attention to couplings $g_0 \leq 1$ later in this paper. As discussed below the effect can be traced back to the occasional presence of exceptionally small eigenvalues of the Dirac operator. The quenched approximation breaks down in this situation, because the contributions of the low-lying modes to fermionic correlation functions such as $f_A$ and $f_P$ are unbounded. In full QCD the singularity is absent and no fundamental difficulty is expected to arise.

### 4.1 Problem description

To illustrate the phenomenon, let us consider the quantity

$$\mathcal{O}(x) = \tfrac{1}{2} \operatorname{tr} \left\{ H(x)^\dagger H(x) \right\}. \tag{4.1}$$

According to eq. (2.7) its average value is equal to $f_P(x_0)$. Most of the time $\mathcal{O}(x)$ is hence fluctuating around $f_P(x_0)$ with some variance $\sigma^2$. The problem occurs when the generated ensemble of gauge field configurations contains a few exceptional configurations, where $\mathcal{O}(x)$ shoots up to values that are orders of magnitude above the normal level of fluctuations characterized by $\sigma$. A reliable estimation of $f_P(x_0)$ is then impossible.

Whether this happens or not depends on $g_0$, $am_q$, $c_{sw}$ and the lattice size $L/a$. In general the fraction of exceptional configurations grows when $g_0$, $c_{sw}$ or $L/a$ is increased or if the quark mass is made smaller. If we choose $\beta = 5.9$, $c_{sw} = 1.6$, $am_q = 0.02$ and $L/a = 8$, for example, one is likely to find an exceptional configuration after every few hundred iterations or so. Large fluctuations have also been observed at $\beta = 6.0$ and $\beta = 6.2$ but only at much smaller quark masses. When $\beta \geq 6.4$ we could set the quark mass to zero or even to small negative values without running into problems. We verified that the effect persists for different choices of the boundary values of the gauge field and on large lattices with physical sizes up to about 2 fm.

## 4.2 Quark zero-modes

If we multiply the linear system (3.2) with $\gamma_5$ so that the hermitean operator

$$Q = \gamma_5(D + \delta D + m_0) \qquad (4.2)$$

appears on the left-hand side, it is immediately clear that a large contribution to $H(x)$ and hence to $\mathcal{O}(x)$ will arise if $Q$ has eigenvalues close to 0. The low-lying eigenvalues of $Q^2$ can be reliably computed using an algorithm described in ref. [18]. It then turns out that the exceptional gauge field configurations are precisely those where $Q^2$ has eigenvalues orders of magnitude smaller than the normal size of the lowest eigenvalue. In other words, the occurrence of near zero-modes is the cause for the observed unbounded fluctuations.

In parameter regions where zero-modes can appear, the quenched approximation for quark correlation functions ceases to be well-defined. This is a fundamental limitation of the quenched approximation and not simply a failure of the simulation algorithm. As already noted above, the probability for large fluctuations increases when the Sheikholeslami-Wohlert term is included in the quark action. The reason for this is currently not known, but it is conceivable that the distribution of the small eigenvalues of the Dirac operator and the violation of chiral symmetry (which is stronger in the unimproved theory) are related to each other.

In physically small volumes the effective gauge coupling is small and perturbation theory may be used to study the spectrum of the Dirac operator. The choice of boundary conditions matters at this point. With Schrödinger functional boundary conditions one finds that the lowest eigenvalue of $Q^2$ never comes close to zero [16,17]. The gap in the spectrum persists even if we set the quark mass $m_q$ to (small) negative values. This is in line with our experience that large fluctuations have not been observed for $\beta \geq 6.4$ and $L/a \leq 16$, which corresponds to physical box sizes $L \leq 0.8$ fm.

Evidently zero-modes are also avoided if the quark mass is positive and not too small. At $\beta = 6.0$, for example, and with $c_{sw}$ as given in sect. 5, quark masses as low as 20 MeV are safe.

### 4.3 QCD with dynamical quarks

The singularity described above is absent in full QCD, because the functional integral over the quark fields is always finite, for any given gauge field configuration and any quark correlation function that one may consider. To show this we substitute $\overline{\psi} \rightarrow \overline{\psi}\gamma_5$ in the functional integral and expand the quark and anti-quark field in eigenmodes of $Q$. The integration then results in a polynomial in the eigenvalues and eigenfunctions of $Q$ and no singularity will ever show up.

While the occurrence of small eigenvalues of the Dirac operator does not present a fundamental problem in full QCD, difficulties can arise when applying the known simulation algorithms. They all follow the same pattern, where one integrates over the quark fields analytically and includes the resulting quark determinant in the simulation algorithm. After generating a representative ensemble of gauge field configurations, one proceeds as in the quenched approximation.

The inclusion of the quark determinant in the simulation algorithm has the effect that gauge fields leading to small eigenvalues of the Dirac operator are generated rarely. But since the quark diagrams one wants to calculate are exceptionally large in this case, one may again end up with statistical fluctuations that are hard to control.

In principle the problem can be solved by splitting the quark determinant into two factors, one of which is incorporated into the simulation algorithm while the other is evaluated together with the quark diagrams. The second factor should be proportional to the product of the lowest few eigenvalues of $Q^2$ in the limit where these tend to zero. As discussed above this ensures that the singularities of the quark diagrams are cancelled. Note that a splitting of the quark determinant of the required type is inherent in the multi-boson algorithm (the current status of the various algorithms has recently been reviewed by Jansen [19]).

# 5. Computation of $c_{sw}$

We now proceed to describe the non-perturbative calculation of the coefficient $c_{sw}$ along the lines explained in sect. 6 of ref. [3]. Throughout this section we set $T = 2L$, $\theta_k = 0$ and

$$(\phi_1, \phi_2, \phi_3) = \tfrac{1}{6}(-\pi, 0, \pi),$$

$$(\phi_1', \phi_2', \phi_3') = \tfrac{1}{6}(-5\pi, 2\pi, 3\pi). \tag{5.1}$$

An important practical criterion for choosing these particular boundary values has been that the induced background field should be weak on the scale of the lattice cutoff to avoid large lattice effects. On the other hand, the effects of order $a$ which one intends to cancel by adjusting $c_{sw}$ should not be too small as otherwise one would be unable to compute $c_{sw}$ accurately. The boundary values (5.1) represent a compromise where both criteria are fulfilled to a satisfactory degree on the accessible lattices.

## 5.1 Improvement condition

As in ref. [3] we introduce an unrenormalized current quark mass $m$ through

$$m = \tfrac{1}{2}\left[\tfrac{1}{2}(\partial_0^* + \partial_0)f_A(x_0) + c_A a \partial_0^* \partial_0 f_P(x_0)\right]/f_P(x_0). \tag{5.2}$$

Another mass, $m'$, may be defined in the same way using the primed correlation functions (cf. sect. 2). The PCAC relation then implies that the mass difference $m - m'$ is of order $a^2$ if the coefficients $c_{sw}$ and $c_A$ are chosen appropriately. Our intention in the following is to take this as a condition to fix $c_{sw}$.

Before being able to do so, we must eliminate the coefficient $c_A$ which is also not known at this point. To this end first note that

$$m(x_0) = r(x_0) + c_A s(x_0), \tag{5.3}$$

where $r$ and $s$ are defined through

$$r(x_0) = \tfrac{1}{4}(\partial_0^* + \partial_0)f_A(x_0)/f_P(x_0), \tag{5.4}$$

$$s(x_0) = \tfrac{1}{2}a\partial_0^* \partial_0 f_P(x_0)/f_P(x_0) \tag{5.5}$$

(for clarity the dependence on the time coordinates is now often indicated explicitly). The other mass $m'$ is similarly given in terms of two ratios $r'$ and $s'$. It is then trivial to show that the combination

$$M(x_0, y_0) = m(x_0) - s(x_0)\frac{m(y_0) - m'(y_0)}{s(y_0) - s'(y_0)} \qquad (5.6)$$

is independent of $c_{\mathrm{A}}$, viz.

$$M(x_0, y_0) = r(x_0) - s(x_0)\frac{r(y_0) - r'(y_0)}{s(y_0) - s'(y_0)}. \qquad (5.7)$$

Furthermore, from eq. (5.6) one infers that $M$ coincides with $m$ up to a small correction of order $a^2$ (in the improved theory). $M$ may hence be taken as an alternative definition of an unrenormalized current quark mass, the advantage being that we do not need to know $c_{\mathrm{A}}$ to be able to calculate it.

We now continue to discuss the condition that determines $c_{\mathrm{sw}}$. If we define $M'$ in the same way as $M$, with the obvious replacements, it follows from the above that the difference

$$\Delta M = M\left(\tfrac{3}{4}T, \tfrac{1}{4}T\right) - M'\left(\tfrac{3}{4}T, \tfrac{1}{4}T\right) \qquad (5.8)$$

must vanish, up to corrections of order $a^2$, if $c_{\mathrm{sw}}$ has the proper value. The coefficient may hence be fixed by calculating $\Delta M$ for a range of values of $c_{\mathrm{sw}}$ and searching for the point where $\Delta M$ passes through zero.

### 5.2 O($a^2$) effects and choice of parameters

As a result of the residual lattice effects of order $a^2$, the values of $c_{\mathrm{sw}}$ calculated in this way are slightly dependent on the quark mass and the lattice size $L/a$. To completely specify the improvement condition, a definite choice of these parameters must be made. It should be obvious that different choices lead to values of $c_{\mathrm{sw}}$ differing by terms of order $a$. Such variations are considered negligible in the O($a$) improved theory, but it is clearly important to check that they are numerically small on the chosen lattices.

In the present context we take $M\left(\tfrac{1}{2}T, \tfrac{1}{4}T\right)$ as the definition of the quark mass. In fact the choice of the quark mass is not critical here since $\Delta M$ is practically independent of $M$ on the level of the statistical errors; a typical case is shown in fig. 1. For definiteness we decided to evaluate $\Delta M$ at $M = 0$ if $\beta \geq 6.4$, while at $\beta = 6.2$ and $\beta = 6.0$ we set $aM = 0.004$ and $aM = 0.01$, respectively. The reason for choosing a non-zero mass at the lower values of $\beta$ is that we would like to avoid the parameter region where quark zero-modes can appear (cf. sect. 4).

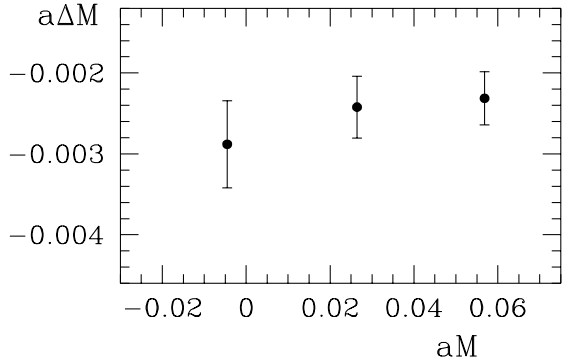

Fig. 1. Mass difference $\Delta M$ on a $16 \times 8^3$ lattice as a function of the quark mass $M$ at $\beta = 6.4$ and $c_{\mathrm{sw}} = 1.777$.

The lattice size $L/a$ should be large so that the higher-order cutoff effects associated with this scale are suppressed. However, when $L/a$ is increased, it becomes more and more difficult to calculate $\Delta M$ accurately. As a result $c_{\mathrm{sw}}$ is obtained with larger statistical errors. A compromise must hence be found where both the statistical errors and the cutoff effects are small.

Some insight into the size of the cutoff effects can be gained in perturbation theory, where

$$\Delta M = \Delta M^{(0)} + g_0^2 \Delta M^{(1)} + \ldots \tag{5.9}$$

Following our discussion above, the expansion is to be performed at a bare quark mass chosen so that $M = 0$ to the order considered. At tree-level we have [4]

$$\Delta M^{(0)} = k(c_{\mathrm{sw}}^{(0)} - 1)a/L + \mathrm{O}(a^2/L^2) \tag{5.10}$$

with some non-zero numerical constant $k$. Imposing $\Delta M = 0$ thus implies $c_{\mathrm{sw}}^{(0)} = 1$ up to a term of order $a/L$. A detailed calculation shows that this correction is only 2% at $L/a = 8$.

In the non-perturbative region the size of the residual cutoff effects has been investigated in several ways. The general idea is to construct alternative improvement conditions and to verify that they give the same result for $c_{\mathrm{sw}}$ within errors. A simple possibility is to replace the 2-point difference $\Delta M$ by a 3-point difference. Other improvement conditions that we have considered are based on the observation that $M$ should be independent of the boundary values of the gauge field up to $O(a^2)$ corrections (in the improved theory). As a result of all these studies we finally decided to take

$$\Delta M = \Delta M^{(0)}|_{M=0,\,c_{sw}=1} \quad \text{at} \quad L/a = 8 \tag{5.11}$$

as the definitive form of the improvement condition for $c_{sw}$. The tree-level term on the right-hand side evaluates to $0.000277/a$ at $L/a = 8$. We have included it for purely aesthetic reasons to cancel the 2% correction mentioned above and thus to guarantee that the non-perturbatively determined $c_{sw}$ is equal to 1 at $g_0 = 0$. The correction is negligible at large bare couplings.

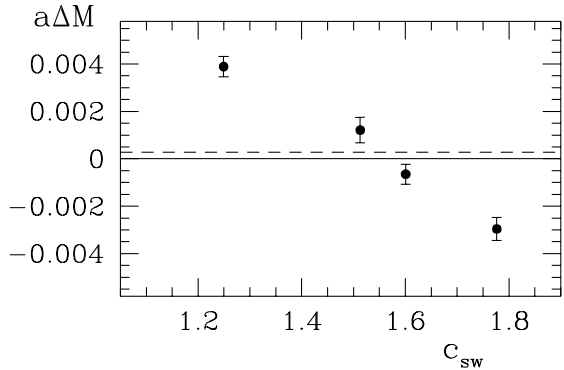

Fig. 2. Determination of $c_{sw}$ at $\beta = 6.4$. The dashed line indicates the tree-level value $\Delta M^{(0)}$ appearing on the right-hand side of the improvement condition (5.11).

### 5.3 Numerical procedure

In practice $c_{sw}$ is obtained by going through the following steps. We first choose the bare coupling $g_0$ at which $c_{sw}$ is to be determined. From perturbation theory or previous non-perturbative calculations of $c_{sw}$ at lower values of $g_0$, it is usually possible to give a rough estimate of $c_{sw}$. We then set $c_{sw}$ to a few values around this estimate and compute $\Delta M$ at all these points.

To calculate $\Delta M$ for any given $c_{sw}$ we choose three to four values of $\kappa$ in the range where $M$ assumes the desired value (zero in most cases). The precise point where this happens and the corresponding value of $\Delta M$ are determined through linear interpolation. Systematic errors from the interpolation can be safely neglected here, because, as noted above, $\Delta M$ is practically independent

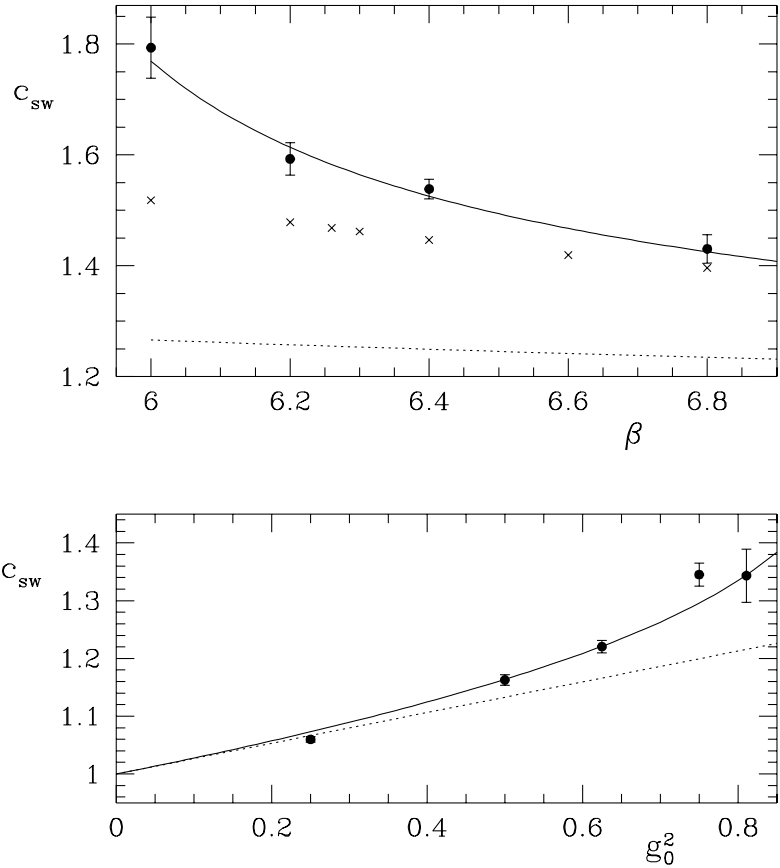

Fig. 3. Results for $c_{sw}$ from numerical simulations (filled circles), bare perturbation theory (dotted line) and "mean field improved" perturbation theory (crosses). The full line represents the fit (5.15).

of $M$ in the interpolation range.

A typical result of such a computation is shown in fig. 2. The available data points are always such that $\Delta M$ is clearly seen to pass through zero. There is no evidence for non-linear behaviour of $\Delta M$ in the intervals of $c_{sw}$ considered. The solution of eq. (5.11) is hence found by linear interpolation.

### 5.4 Results

The non-perturbative calculation of $c_{\text{sw}}$ along the lines explained above has been carried out at 9 values of the bare coupling. The results are shown in fig. 3. In the range of couplings relevant for numerical simulations in physically large volumes, $c_{\text{sw}}$ is substantially larger than the tree-level value $c_{\text{sw}} = 1$. It is reassuring, however, that the one-loop formula [20,21,4]

$$c_{\text{sw}} = 1 + c_{\text{sw}}^{(1)} g_0^2 + \mathrm{O}(g_0^4), \qquad c_{\text{sw}}^{(1)} = 0.2659(1), \tag{5.12}$$

describes the data rather well for say $g_0^2 \leq 0.5$.

As discussed in ref. [22] mean field theory suggests to rewrite eq. (5.12) in the form

$$c_{\text{sw}} = u_0^{-3} \left[ 1 + \left( c_{\text{sw}}^{(1)} - 1/4 \right) g_{\text{P}}^2 + \mathrm{O}(g_{\text{P}}^4) \right], \tag{5.13}$$

where $u_0^4$ is the average plaquette in infinite volume at the value of $g_0^2$ considered and [23]

$$g_{\text{P}}^2 = g_0^2 / u_0^4. \tag{5.14}$$

The "mean field improved" formula in fact comes much closer to the data than bare perturbation theory (see fig. 3). A discrepancy of up to 20% however remains at low values of $\beta$.

Our numerical results are well represented by the rational expression

$$c_{\text{sw}} = \frac{1 - 0.656\, g_0^2 - 0.152\, g_0^4 - 0.054\, g_0^6}{1 - 0.922\, g_0^2}, \qquad 0 \leq g_0 \leq 1. \tag{5.15}$$

The precision of this parametrization is around 3% in the whole range of couplings. It has been chosen so that the one-loop formula (5.12) is reproduced at small $g_0$. In the following we *define* $c_{\text{sw}}$ through eq. (5.15). From the point of view of improvement this is as good as taking the numerically determined values of $c_{\text{sw}}$. We however prefer to have a lattice action which is defined without numerical uncertainty. The hadron masses and other quantities that one may wish to calculate are then guaranteed to be smooth functions of the bare coupling. Extrapolations to the continuum limit would otherwise be quite impossible.

## 6. Computation of $c_A$

Now that the improved action is known, it is relatively easy to calculate $c_A$. An important qualitative observation is that the cutoff effects which one intends to cancel by adjusting $c_A$ are rather small in general. The calculated values of $c_A$ are hence more sensitively dependent on the chosen improvement condition than in the case of $c_{sw}$. As already emphasized in subsect. 6.5 of ref. [3], there is nothing fundamentally wrong with this. Our aim is to minimize the cutoff effects by tuning $c_A$, and this can be achieved by any careful choice of the improvement condition.

### 6.1 Improvement condition

In the following we again set $T = 2L$ but choose the boundary values $C$ and $C'$ of the gauge field to vanish. Perturbation theory [4] and some preliminary studies made in the course of the calculations described in sect. 5 suggest that this is a good choice for the determination of $c_A$. As before we define the unrenormalized current quark mass $m$ through eq. (5.2). $m$ is always evaluated at time $x_0 = T/2$ unless specified otherwise.

The improvement condition for $c_A$ that we have chosen is based on the observation that the mass difference

$$\Delta m = m|_{\theta_k=1} - m|_{\theta_k=0} \tag{6.1}$$

must vanish in the improved theory up to terms of order $a^2$ (cf. sect. 6 of ref. [3]). Explicitly $c_A$ is determined by requiring

$$\Delta m = \Delta m^{(0)}|_{m=0, c_A=0} \quad \text{at} \quad L/a = 8, \tag{6.2}$$

where $\Delta m^{(0)}$ denotes the tree-level value of $\Delta m$. Eq. (6.2) is to be evaluated at quark mass $m|_{\theta_k=0} = 0$ if $\beta \geq 6.2$, while at $\beta = 6.1$ and $\beta = 6.0$ we set $am = 0.008$ and $am = 0.023$, respectively. These choices are not critical since $\Delta m$ is only weakly dependent on the quark mass (see fig. 4).

To lowest order of perturbation theory the condition $\Delta m = 0$ would imply a non-zero value of $c_A$ of order $a/L$. The tree-level correction on the right-hand side of eq. (6.2) has been included to cancel this effect and thus to enforce that the non-perturbatively determined $c_A$ vanishes at $g_0 = 0$ [4,24]. The correction is numerically small,

$$\Delta m^{(0)}|_{m=0, c_A=0} = 0.000365/a \quad \text{at} \quad L/a = 8, \tag{6.3}$$

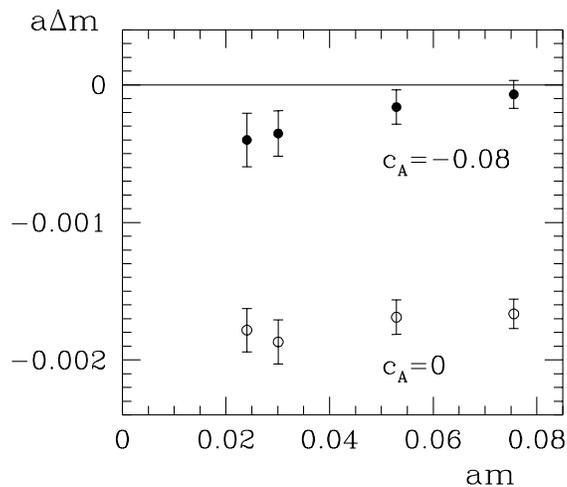

Fig. 4. Mass difference $\Delta m$ on a $16 \times 8^3$ lattice as a function of the quark mass $m$ at $\beta = 6.0$ and two values of $c_A$.

and does not have a strong influence on the calculated values of $c_A$ at larger couplings.

As in the case of the coefficient $c_{sw}$, the lattice size, $L/a = 8$, has been chosen after performing a number of checks on the magnitude of residual cutoff effects of order $a^2$. In particular we set $c_A$ to the value determined through eq. (6.2) and then investigated the dependence of the current quark mass $m$ on the position $x_0$ for both $\theta = 0$ and $\theta = 1$. For $\beta \geq 6.4$, varying $x_0$ from $T/4$ to $3T/4$ the change of $am$ is found to be below the $10^{-3}$–level, i.e. below the expected order of $(a/L)^3$. On the other hand at the lowest values of $\beta$ non-perturbative higher order cutoff effects are visible. An example of the latter will be discussed in more detail in section 7.2.

### 6.2 Numerical procedure and results

The quark mass $m$ and the mass difference $\Delta m$ are linearly dependent on $c_A$. Our task is to solve eq. (6.2) for $c_A$ at a specified value of $m$. This is achieved by calculating the required ratios $r$ and $s$ introduced in subsect. 5.1 for a few values of $\kappa$ in the relevant range. At each point we determine $c_A$ from eq. (6.2) and use this number to calculate the quark mass $m$. The value of $c_A$ at the desired quark mass $m$ is then found by linear interpolation.

The result of our calculations is shown in fig. 5. As already suspected from perturbation theory [4], $c_{\rm A}$ is rather small and remains so even at the largest values of the bare coupling considered. The one-loop formula,

$$c_{\rm A} = c_{\rm A}^{(1)} g_0^2 + {\rm O}(g_0^4), \qquad c_{\rm A}^{(1)} = -0.00756(1), \qquad (6.4)$$

describes the data rather well for $g_0^2 \leq 0.5$, thus giving further evidence for the smallness of the residual cutoff effects in our determination of $c_{\rm A}$. "Mean field improved" perturbation theory here amounts to replacing $g_0^2$ by $g_{\rm P}^2$ [cf. eq. (5.14)], but as can be seen from fig. 5 this modification cannot make up for the large difference between perturbation theory and the data at low values of $\beta$.

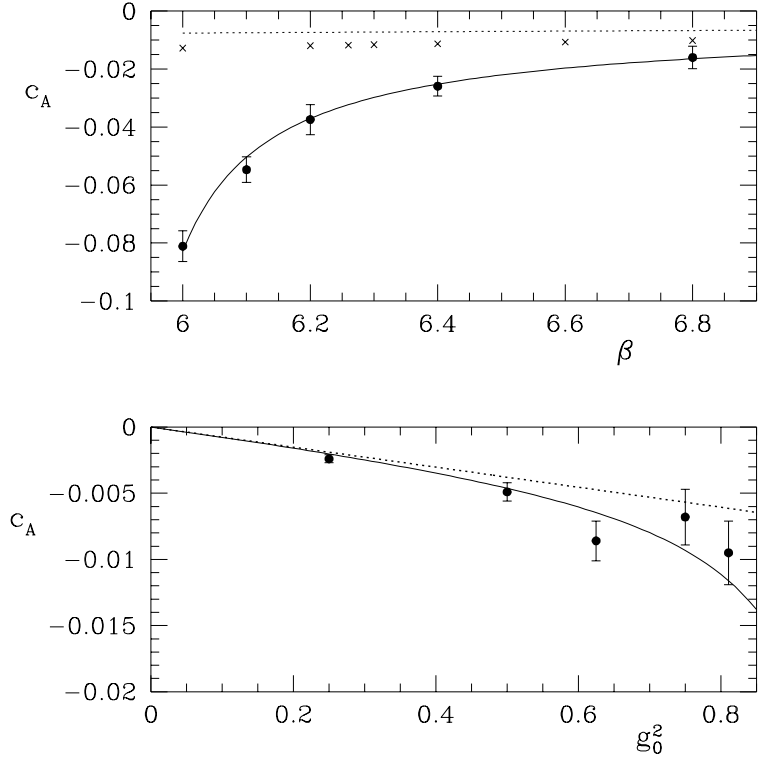

Fig. 5. Results for $c_{\rm A}$ from numerical simulations (filled circles), bare perturbation theory (dotted line) and "mean field improved" perturbation theory (crosses). The full line represents the fit (6.5).

For future applications it is again convenient to represent our results in the form of an approximate analytic expression. A good representation of the data shown in fig. 5, which incorporates the required asymptotic behaviour (6.4) at small couplings, is given by

$$c_A = -0.00756 \, g_0^2 \times \frac{1 - 0.748 \, g_0^2}{1 - 0.977 \, g_0^2}, \qquad 0 \leq g_0 \leq 1. \tag{6.5}$$

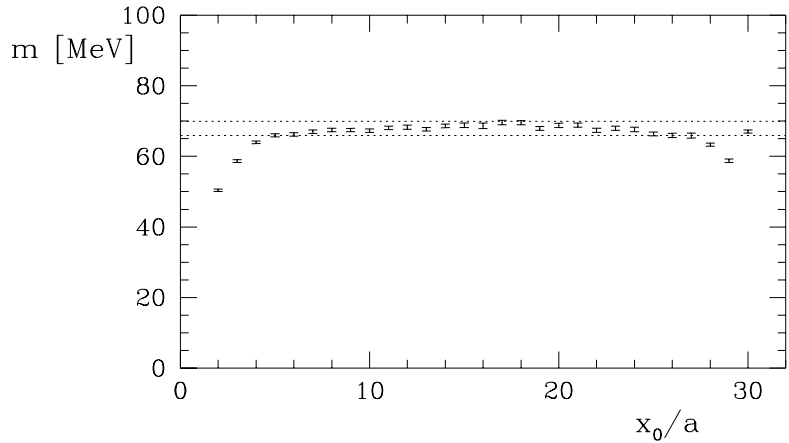

Fig. 6. Unrenormalized current quark mass $m$ in the improved theory, with non-perturbatively determined $c_{sw}$ and $c_A$, as a function of the time $x_0$ on a $32 \times 16^3$ lattice at $\beta = 6.2$ and $\kappa = 0.1350$. The width of the corridor bounded by the dotted horizontal lines is $4 \, \mathrm{MeV}$.

We finally remark that the PCAC relation is accurately satisfied once improvement has been fully implemented. To illustrate this we plot the quark mass $m$ against the time $x_0$ on a large lattice at $\beta = 6.2$ (see fig. 6). Here $m$ is given in physical units using the radius $r_0$ [25,26] to set the scale. The figure shows that the data are nearly independent of $x_0$ for $5 \leq x_0/a \leq 27$. The residual cutoff effects are on the level of $\pm 2 \, \mathrm{MeV}$ in this range. Away from the boundaries of the lattice the violations of the PCAC relation in the improved theory are hence rather small.

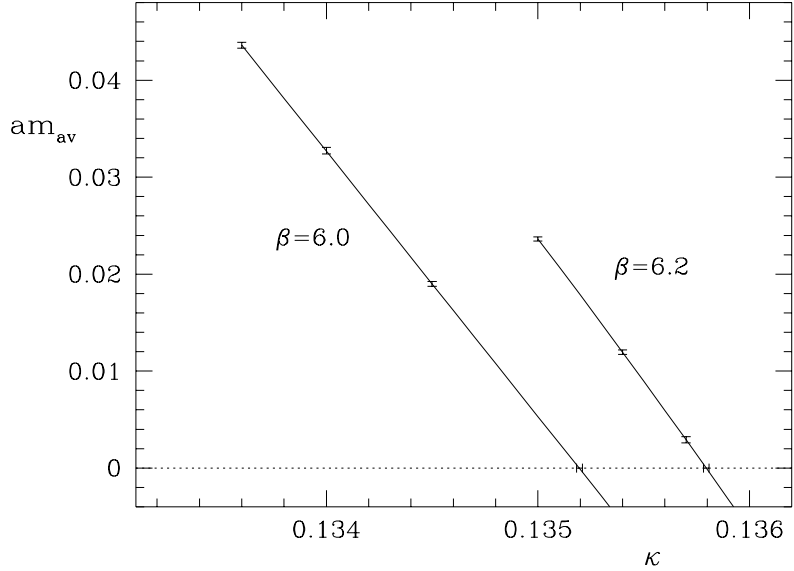

Fig. 7. Extrapolation of the data for $am_{av}$ at $\beta = 6.0$ and $\beta = 6.2$. The points with the horizontal error bars represent the calculated values of $\kappa_c$.

## 7. The critical line

The critical hopping parameter $\kappa_c$ is here defined to be the value of $\kappa$ where the unrenormalized current quark mass $m$ vanishes. As already pointed out in subsect. 6.6 of ref. [3], other definitions are possible and this leads to a systematic uncertainty in the calculated values of $\kappa_c$ which is of order $a^3$ in the improved theory. We now first describe our computation of $\kappa_c$ in some detail and resume the discussion of cutoff effects in subsect. 7.2. In all calculations reported in this section the analytic expressions (5.15) and (6.5) have been used for $c_{sw}$ and $c_A$. The boundary values of the gauge field and the angles $\theta_k$ are set to zero and we always take $T = 2L$ as before.

### 7.1 Computation of $\kappa_c$

For reasons to be given later we decided to calculate the critical hopping parameter using a lattice of size $L/a = 16$. As shown in fig. 6 the quark mass $m$ is then nearly independent of $x_0$ around $x_0 = T/2$. Somewhat surprisingly it turns out that the statistical errors of the data at different times are not

Table 1. Values of the critical hopping parameter

| $\beta$ | $\kappa_c$ | $\beta$ | $\kappa_c$ |
|------|-----------|------|-----------|
| 6.0 | 0.135196(14) | 8.0 | 0.133173(3) |
| 6.2 | 0.135795(13) | 9.6 | 0.131448(2) |
| 6.4 | 0.135720(9) | 12.0 | 0.129909(2) |
| 6.8 | 0.135097(5) | 24.0 | 0.127258(1) |
| 7.4 | 0.134071(4) | | |

very strongly correlated so that the signal-to-noise ratio can be enhanced by taking the average

$$m_{\mathrm{av}} = \tfrac{1}{5} \sum_{x_0=14a}^{18a} m(x_0). \tag{7.1}$$

We then solve the equation $m_{\mathrm{av}} = 0$ for $\kappa_c$ by calculating $m_{\mathrm{av}}$ at two values of $\kappa$, one slightly above $\kappa_c$ and the other slightly below, and subsequent linear interpolation.

A different procedure has to be applied at $\beta = 6.2$ and $\beta = 6.0$, where simulation results for $m_{\mathrm{av}}$ are only available at values of $\kappa$ below $\kappa_c$. In the first case a quadratic extrapolation is used to determine $\kappa_c$ (see fig. 7). Linear extrapolation yields compatible results. At $\beta = 6.0$ the data points closest to $\kappa_c$ are extrapolated linearly. Including additional points in the fit, at quark masses around 0.065 and 0.087, gives results for $\kappa_c$ well within the quoted error margin.

Our results for $\kappa_c$ at various bare couplings are listed in table 1. In contrast to the unimproved theory, the numbers found here are rather close to the tree-level value 0.125. One may thus be led to expect that the one-loop formula [21,27,4]

$$\kappa_c = \tfrac{1}{8} + \kappa_c^{(1)} g_0^2 + \mathrm{O}(g_0^4), \qquad \kappa_c^{(1)} = 0.00843986(4), \tag{7.2}$$

gives a good description of the data. This is not the case, however, and the "mean field improved" formula [22],

$$\kappa_c = \frac{1}{8u_0} \big[ 1 + \big( 8\kappa_c^{(1)} - 1/12 \big) g_{\mathrm{P}}^2 + \mathrm{O}(g_{\mathrm{P}}^4) \big], \tag{7.3}$$

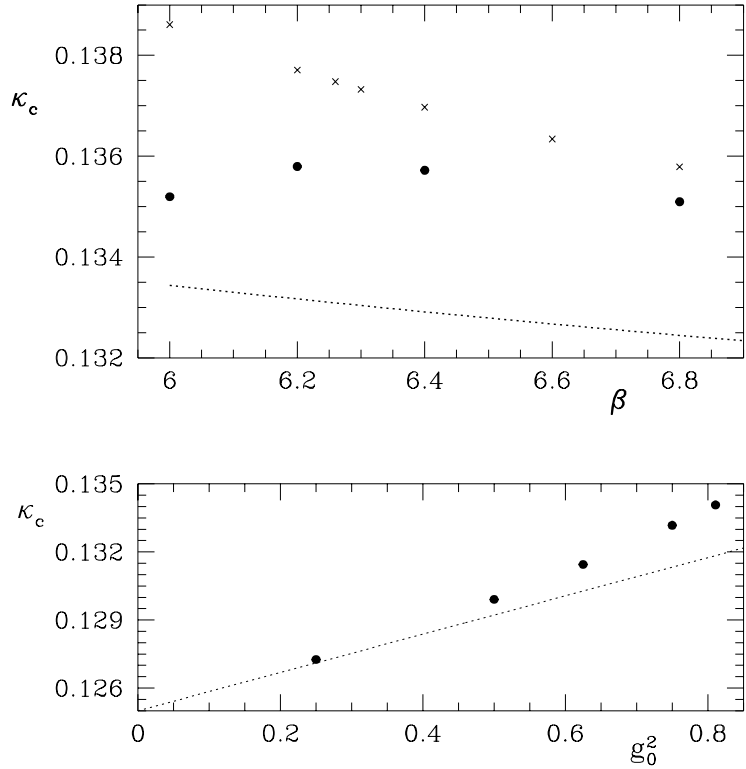

Fig. 8. Results for $\kappa_c$ from numerical simulations (filled circles), bare perturbation theory (dotted line) and "mean field improved" perturbation theory (crosses). The statistical errors are not visible on this scale.

does not fare much better at low values of $\beta$ (see fig. 8).

### 7.2 Systematic uncertainties in $\kappa_c$

So far we have set $L/a = 16$ and we would now like to justify this choice and to obtain some insight into the size of the associated systematic uncertainty in $\kappa_c$.

We first discuss the issue in perturbation theory, where the cutoff effects on the calculated values of $\kappa_c$ are of order $(a/L)^3$. From the one-loop results listed in sect. 7 of ref. [4] one expects to see variations in $\kappa_c$ of at most $2 \times 10^{-5}$ when $L/a$ is increased from 8 to 16. The statistical errors quoted in table 1 are of the same order of magnitude at the lower values of $\beta$. It is essentially for this reason that we have decided to take $L/a = 16$ in our calculations of $\kappa_c$. The systematic uncertainty is then reduced by a factor 8 to a level well below $10^{-5}$.

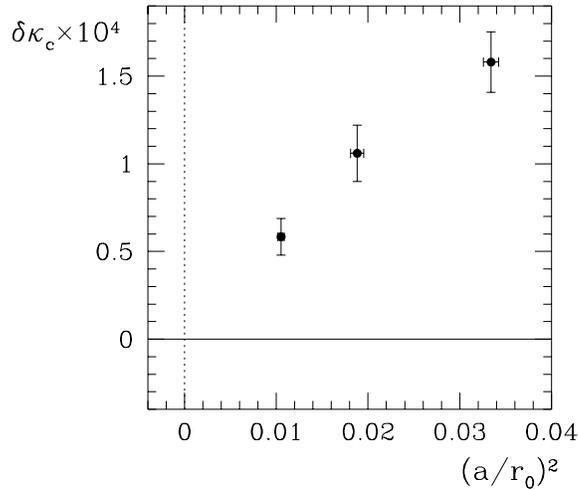

Fig. 9.   Change $\delta\kappa_c$ in the calculated value of $\kappa_c$ when $L/a$ is increased from 8 to 16. The data have been obtained at $\beta = 6.0, 6.2$ and $6.4$ (from right to left).

In the non-perturbative region the situation is more complicated, because one also has the dynamical length scales, such as the radius $r_0$ extracted from the force between heavy quarks [25]. In particular, cutoff effects of order $a^3$ are equal to $(a/L)^3$ times an unknown function of $L/r_0$. As a consequence the systematic uncertainties on $\kappa_c$ may be larger than suggested by perturbation theory and they are not necessarily decreasing proportionally to $(a/L)^3$ at fixed bare coupling.

To study this question we compare the results listed in table 1 with the values of $\kappa_c$ that were obtained in the course of the calculation of $c_A$. In sect. 6 the zero-mass point has been defined exactly as above, the only difference being that a lattice of size $L/a = 8$ has been employed. It turns out that the calculated values of $\kappa_c$ agree to an accuracy better than $2 \times 10^{-5}$ at all couplings $\beta \geq 6.8$. In this range there is hence no sign of a non-perturbative cutoff effect and it appears safe to say that our results for $\kappa_c$ would change by no more than $10^{-5}$ if the lattice size $L/a$ would be increased to values greater than 16.

As shown in fig. 9 a significant shift in $\kappa_c$ is however found at the lower values of $\beta$, where $r_0/a$ varies between 5 and 10 [26]. The data points plotted in the figure suggest that the dominant effect is proportional to $(a/L)(a/r_0)^2$, but much more detailed work would be needed to firmly establish this. In any case, our results clearly show that non-perturbative residual cutoff effects are present and that they can be quite significant at low values of $\beta$.

## 8. Concluding remarks

For future work in quenched $O(a)$ improved lattice QCD we propose to define the coefficients $c_{sw}$ and $c_A$ through eqs. (5.15) and (6.5). This guarantees an almost perfect cancellation of the $O(a)$ lattice corrections in spectral quantities and on-shell matrix elements of the axial current and density. Moreover, as we have shown in a number of cases, the remaining higher-order lattice corrections to the PCAC relation are small at all values of the bare coupling considered. For very accurate results, and to control the systematic uncertainties associated with the residual cutoff effects, an extrapolation to the continuum limit is however still required.

It is obviously important to study the hadron spectrum and hadronic matrix elements in the improved theory with the non-perturbatively determined $c_{sw}$ and $c_A$. First steps in this direction have already been taken [28,29] and it is clear from these calculations that improvement has a significant impact on the hadron masses and decay constants. One would now need to check that the continuum limit is indeed reached more rapidly than in the unimproved theory.

It has recently been shown [30] that the overhead associated with the Sheikholeslami-Wohlert term is small compared to the total cost of simulations of lattice QCD with dynamical quarks. Since we did not need to simulate very large lattices to calculate $c_{sw}$ and $c_A$, we are confident that the computations can be extended to full QCD with say two flavours of light quarks. The calculation has in fact already been initiated and the experience made so far confirms our expectations.

This work is part of the ALPHA collaboration research programme. We thank DESY for allocating computer time on the APE/Quadrics computers at DESY-IfH and the staff of the computer center at Zeuthen for their support. We are grateful to Karl Jansen for helpful conversations and to Hartmut Wittig for a critical reading of a first draft of the paper. Stefan Sint is partially sup-